\documentclass[a4paper,floatfix,amsmath,amssymb,pre,showpacs,twocolumn,superscriptaddress,nofootinbib]{revtex4-1}
\usepackage{amsmath}
\usepackage{epsfig}
\usepackage{epstopdf}
\usepackage{graphicx}% include figure files
\usepackage{epstopdf}
\usepackage{csquotes}
\usepackage{subfig}
\usepackage{hyperref}
\hypersetup{
pdfstartview=FitV,
colorlinks=true, 
linkcolor=blue, 
citecolor=blue, 
filecolor=black, 
urlcolor=blue,
pdftitle={Feature-rich bifurcations in a simple electronic circuit}}
\usepackage{dcolumn}% align table columns on decimal point
\usepackage{bm}% bold math
\usepackage{color}% color

\newcommand{\newc}{\newcommand}
\newc{\mbf}{\mathbf}
\newc{\boma}{\boldmath}
\newc{\phihat}{\mbox{\boldmath{$\hat{\phi}$}}}
\newc{\thetahat}{\mbox{\boldmath{$\hat{\theta}$}}}
\newc{\beq}{\begin{equation}}
\newc{\eeq}{\end{equation}}
\newc{\beqar}{\begin{eqnarray}}
\newc{\eeqar}{\end{eqnarray}}
\newc{\beqa}{\begin{eqnarray*}}
\newc{\eeqa}{\end{eqnarray*}}
\newc{\bd}{\begin{displaymath}}
\newc{\ed}{\end{displaymath}}
\begin{document}

\title{Feature-rich bifurcations in a simple electronic circuit}

\author{Debdipta Goswami} %$^1$, J. Shamanna$^2$ and Subhankar Ray$^3$}
\email{goswamid@umd.edu}
\affiliation{Department of Electrical and Computer Engineering,\\
University of Maryland, College Park, MD 20742, USA}
\author{Subhankar Ray}
\email{sray.ju@gmail.com}
\affiliation{Department of Physics, Jadavpur University, 
Calcutta 700 032, India}

%\pacs{81.05.Rm, 68..35.-p, 68.35.Ja}
%\date{\today}{16 April 2017}
\date{16 April 2017}

\begin{abstract}
\noindent A simple electronic circuit with a voltage controlled current source is investigated. The circuit exhibits rich dynamics upon varying the circuit elements such as L,C and R, and the control factor of the current source. Among several other interesting features, the circuit demonstrates two local bifurcations, namely, node to spiral and Hopf bifurcation, and a global homoclinic bifurcation.
Phase-portraits corresponding to these bifurcations are presented and the implications of these
bifurcations on system stability are discussed. In particular, the circuit parameters corresponding
to the onset of Hopf bifurcation may be exploited to design an oscillator with stable frequency and amplitude. The circuit may be easily implemented with nonlinear resistive elements such as diodes or transistors in saturation and a gyrator block as the voltage controlled current source.
\end{abstract}

\keywords{Electronic circuit model ; bifurcations; nonlinear differential equations; dynamical
system; spiral-node bifurcation; Hopf bifurcation; homoclinic bifurcation.}
%\maketitle   -for separate title page
\maketitle

\section{Introduction}
\label{intro}
The theory of dynamical systems and the underlying miracles of bifurcations are wisely used in the fields of theoretical sciences and engineering\cite{ref1}-\cite{ref7}. In the domain of electronics, from Chua's circuit  to Vander-pole oscillator, whenever a system is modelled by a set of differential equations, the study of bifurcations has strengthened the theoretical basis of a particular circuit model and at the same time it often opens new portals to synthesis of more accurate and stable circuits. Knowing the bifurcation-point is very important because it denotes a transition from one regime of dynamics to another.

Bifurcations can be local or global ones. Bifurcations like Hopf or saddle-node are of the former category as they can be detected by local analysis of a fixed point\cite{ref7}. But the bifurcations like Homoclinic where a periodic solution collides with a steady-state one cannot be detected by local analysis and thus called global bifurcations.
 
For a long time researchers have been attracted by the non-linearity in a system that can be properly exploited by tuning the system parameters so that it results in a stable and controllable desired behaviour. We may take the example of Vander-pole oscillator where, the intrinsic non-linearity of the system helps to stabilize the amplitude of the oscillation which cannot be done in a simple phase-shift or LC oscillator without an amplitude control mechanism\cite{refvanderpol1},\cite{refvanderpol2}. Practical dynamical systems that are observed in nature by means of population drift or fluid mechanics can easily be modelled and analyzed by electrical analog circuits where the different decisive parameters can be controlled as per the wish of the researcher to gain a better insight\cite{ref5}, \cite{ref6}. Often exploiting those dynamics on circuitry results in the discovery of a new stable circuit that can be extensively used in growing electronics industry.

Thus, keeping in mind the wide applicability and the intrinsic mathematical interest, I propose in this article a simple electrical circuit model that displays a feature-rich dynamics with different local and global bifurcations. The bifurcations are mathematically analyzed and experimentally simulated. The parameter for homoclinic bifurcation which cannot be determined by direct analytical method is determined from the quantitative details that the experiments provide. The phase-portraits illustrating the bifurcations and different regions of parameter space is also presented.

\section{The Circuit Model}

The proposed circuit model is given in the figure 1. It is implemented using two infinite-gain operational amplifiers. There are two voltage controlled current sources (VCCS) which can be implemented by gyrator blocks. The boxed element is a current squaring device which can be viewed as a non-linear resistor with characteristics as $v=ki^2$. This can be implemented by squaring circuit available in the literature\cite{ref9}, \cite{ref10}. These squarers mainly use MOS circuitry under saturation region. Such a circuit proposed by Sakul\cite{ref10} is shown in the figure 2. Other than these the circuit elements are passive and linear like resistors and inductors. 

\begin{figure}[t]
  %\begin{center}
   %\includegraphics{Z_x_1.jpg}
 \includegraphics[trim=0cm 0cm 0cm 0cm, clip=true, height=7cm, width=0.5 \textwidth]{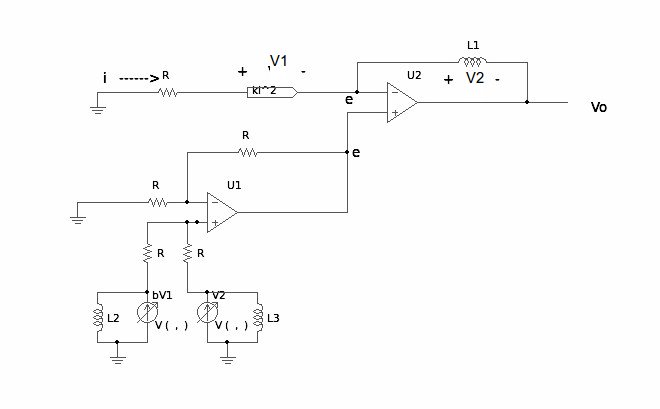}  
  \caption{Circuit Model}
 %\end{center}
\end{figure}

\begin{figure}[h]
  %\begin{center}
   %\includegraphics{Z_x_1.jpg}
\includegraphics[trim=0cm 0cm 0cm 0cm, clip=true, height=7cm, width=0.3 \textwidth]{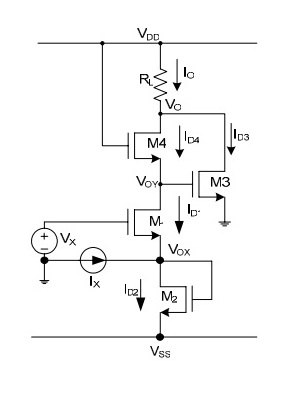}  
  \caption{A CMOS Voltage and Current Squaring Circuit}
 %\end{center}
\end{figure}

The magnitudes of the VCCS elements are given by 
\begin{equation}
bv_1=bki^2
\end{equation}
and
\begin{equation}
v_2=bL_1\dfrac{di}{dt}.
\end{equation}

The lower op-amp acts as an non inverting summer producing the output 
\[e=-(bL_2\dfrac{dv_1}{dt}+L_3\dfrac{dv_2}{dt})\]
\[=-(2bkL_2i\dfrac{di}{dt}+L_1L_3\dfrac{d^2i}{dt^2}).\]

The Kirchoff's voltage law for the branch current $i$ is given by,
\[e=-iR-ki^2.\]

Comparing these two equations we get,
\begin{equation}
iR+ki^2=2bkL_2i\dfrac{di}{dt}+L_1L_3\dfrac{d^2i}{dt^2};
\end{equation}
which ultimately reduces to 
\begin{equation}
\label{circuit_eqn}
\dfrac{d^2i}{dt^2}=-\gamma i\dfrac{di}{dt}+\alpha i+\beta i^2 ,
\end{equation}
where $\alpha=\dfrac{R}{L_1L_3}, \beta=\dfrac{k}{L_1L_3}$ and $\gamma=-\dfrac{2bkL_2}{L_1L_3}$.

Here $\gamma$ is a controllable parameter and can be both positive and negative since the control factor $b$ of the VCCS $bv_1$ can be positive as well as negative.

Now the equation~\ref{circuit_eqn} can be written as a system of differential equations as,
%\begin{subequations}

\begin{equation}
\label{dynamics}
\begin{split}
\dot{x} = y \\
%\end{equation}
%\begin{equation}
\dot{y} = \alpha x +\beta x^2-\gamma xy
\end{split}
\end{equation}
%\end{subequations}

\section{Bifurcations and Phase Portraits}
The system of equations ~\ref{dynamics} will have an equilibrium point $(x^*,y^*)$ when both the derivatives are zero; i.e.
\begin{equation}
\begin{split}
y^*=0\\\alpha x^* +\beta {x^*}^2-\gamma x^*y^*=0.
\end{split}
\end{equation}

\begin{figure}[h]
  \centering
 \includegraphics[trim=0cm 0cm 0cm 0cm, clip=true, height=5cm, width=0.5 \textwidth]{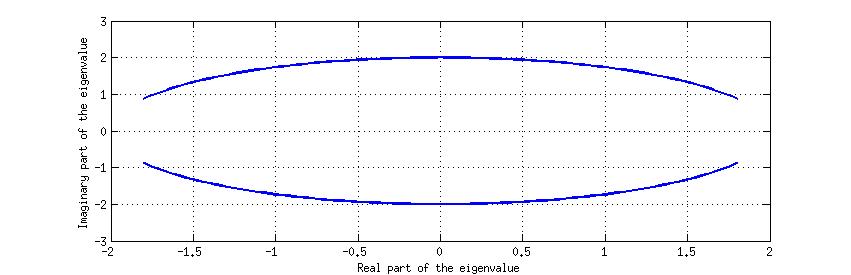}  
  \caption{Eigenvalues of J with the change in $\gamma$}
 %\end{center}
\end{figure}

The solutions of this system is given by $(x^*, y^*)=(0,0)$ and $(x^*, y^*)=(-\dfrac{\alpha}{\beta},0)$.

\begin{figure}[h]
  \centering
 \includegraphics[trim=0cm 0cm 0cm 0cm, clip=true, height=5cm, width=0.4 \textwidth]{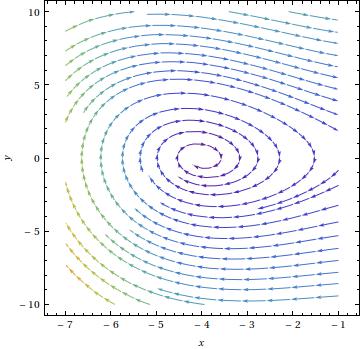}  
  \caption{weak stable spiral at $\gamma=-0.1$}
 %\end{center}
\end{figure}

\begin{figure}[h]
  \centering
 \includegraphics[trim=0cm 0cm 0cm 0cm, clip=true, height=5cm, width=0.4 \textwidth]{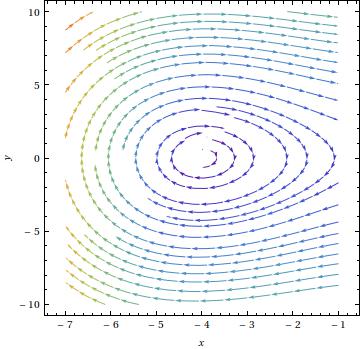}  
  \caption{unstable spiral with limit cycle at $\gamma=+0.05$}
 %\end{center}
\end{figure}

If we linearize the governing equations about these fixed point, we get Jacobian matrix as,
 \begin{equation}
 J=\begin{bmatrix}
 0 & 1 \\ \alpha+2\beta x^*-\gamma y^* & -\gamma x^*
 \end{bmatrix}
 \end{equation}

At $(x^*, y^*)=(0,0)$, $Tr(J)=0$ and $Det(J)=-\alpha$. Since $\alpha>0$, this fixed point is a saddle for the entire parameter space. Now for $(x^*, y^*)=(-\dfrac{\alpha}{\beta},0)$, we get $Tr(J)=\dfrac{\gamma\alpha}{\beta}$ and $Det(J)=\alpha$. So the corresponding eigenvalues are given as 
\begin{equation}
\lambda_{1,2}=\dfrac{1}{2}\biggl[\dfrac{\gamma\alpha}{\beta}\pm\sqrt{\biggl(\dfrac{\gamma\alpha}{\beta}\biggr)^2-4\alpha}\biggr]
\end{equation}

\begin{figure}[h]
  \centering
 \includegraphics[trim=0cm 0cm 0cm 0cm, clip=true, height=4.5cm, width=0.4 \textwidth]{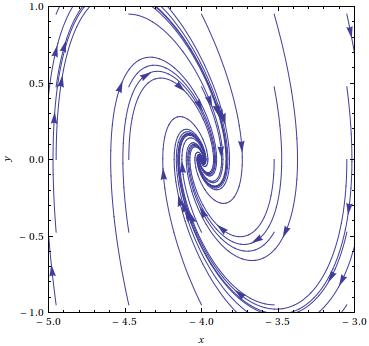}  
  \caption{stable spiral at $\gamma=-0.5$}
 %\end{center}
\end{figure}

\begin{figure}[h]
  \centering
 \includegraphics[trim=0cm 0cm 0cm 0cm, clip=true, height=4.5cm, width=0.4 \textwidth]{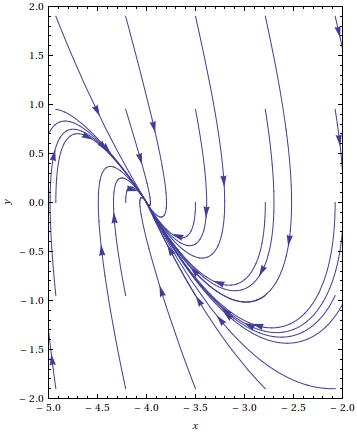}  
  \caption{Degenerate node at $\gamma=-1.0$}
 %\end{center}
\end{figure}

Here it is seen that, since $\alpha$ and $\beta$ are both positive for entire parameter space, the real part of the eigenvalues will change sign from negative to positive whenever $\gamma$ has a similar change of sign. Again when $\biggl(\dfrac{\gamma\alpha}{\beta}\biggr)^2<4\alpha$, i.e. $-\dfrac{2\beta}{\alpha}<\gamma<\dfrac{2\beta}{\alpha}$, the eigenvalues are complex conjugate in nature.
\begin{figure}[h]
  %\begin{center}
   %\includegraphics{Z_x_1.jpg}
 \includegraphics[trim=0cm 0cm 0cm 0cm, clip=true, height=5cm, width=0.4 \textwidth]{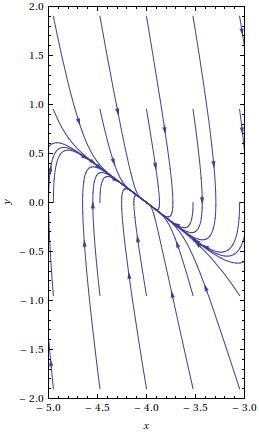}  
  \caption{Stable node with $\gamma=-1.5$}
 %\end{center}
 \end{figure} 
 
 \begin{figure}[h]
  \centering
 \includegraphics[trim=0cm 0cm 0cm 0cm, clip=true, height=7cm, width=0.5 \textwidth]{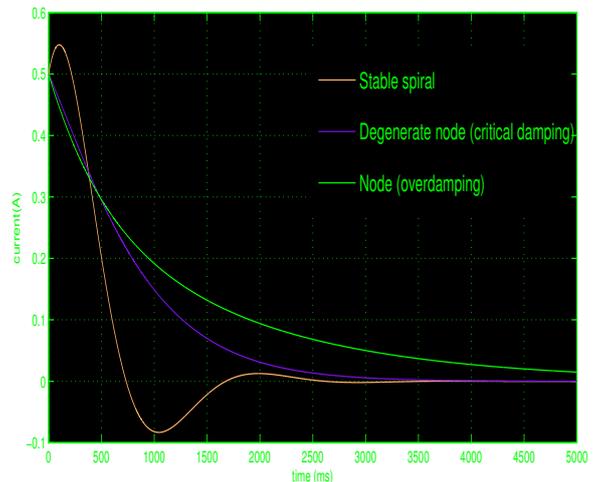}  
  \caption{Time response for stable spiral, stable degenerate node and stable node}
 %\end{center}
\end{figure}

So when $\gamma$ changes its sign from negative to positive, the eigenvalues cross the imaginary axis from left half to right half of the complex plane. So at this point a supercritical Hopf bifurcation will occur. The behaviour of the eigenvalues is shown in the figure 3.

In this bifurcation a stable spiral changes into an unstable spiral surrounded by a nearly elliptical limit cycle. The phase portrait describing this is shown in figure 4 and 5 respectively.

\begin{figure}[h]
  \centering
 \includegraphics[trim=0cm 0cm 0cm 0cm, clip=true, height=5cm, width=0.4 \textwidth]{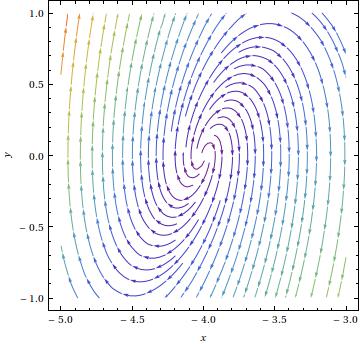}  
  \caption{Unstable spiral at $\gamma=0.5$}
 %\end{center}
\end{figure}

\begin{figure}[h]
  \centering
 \includegraphics[trim=0cm 0cm 0cm 0cm, clip=true, height=5cm, width=0.4 \textwidth]{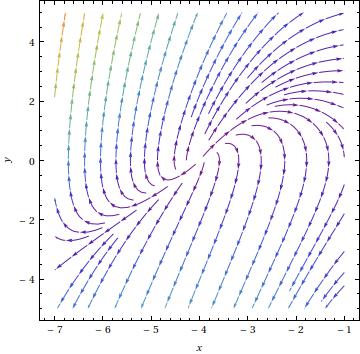}  
  \caption{Unstable degenerate node at $\gamma=1.0$}
 %\end{center}
\end{figure}

\begin{figure}[h]
  \centering
 \includegraphics[trim=0cm 0cm 0cm 0cm, clip=true, height=5cm, width=0.4 \textwidth]{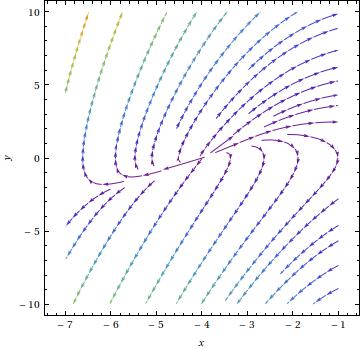}  
  \caption{Unstable node with $\gamma=1.2$}
 %\end{center}
\end{figure}

\begin{figure}[h]
\centering
  %\begin{center}
   %\includegraphics{Z_x_1.jpg}
 \includegraphics[trim=0cm 0cm 0cm 0cm, clip=true, height=5cm, width=0.4 \textwidth]{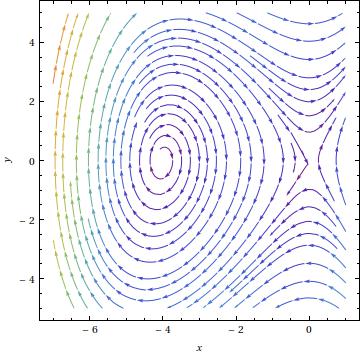}  
  \caption{Homoclinic orbit at $\gamma\approx0.10$}
 %\end{center}
\end{figure}

\begin{figure}[b]
\centering
  %\begin{center}
   %\includegraphics{Z_x_1.jpg}
 \includegraphics[trim=0cm 0cm 0cm 0cm, clip=true, height=5cm, width=0.4 \textwidth]{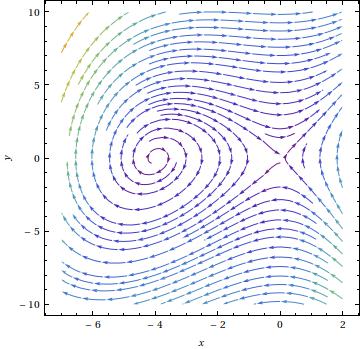}  
  \caption{Unstable spiral and saddle at $\gamma=0.18$}
 %\end{center}
\end{figure}

Another type of local bifurcation can be observed when the eigenvalues become real from complex conjugate and the phase portrait changes from a spiral to node through a degenerate node. When $\gamma$ crosses the value of $-\dfrac{2\beta}{\sqrt{\alpha}}$ and becomes less than it, the fixed point becomes a stable node from a stable spiral. This is shown on figures 6,7 and 8.

The time response for stable spiral degenerate node and node are underdamped, critically damped and overdamped ones, and are shown in figures 9. When we need a noisy oscillation to die down, degenerate node, i.e. critical damping is important because it dies down faster than the other two.

Similarly when $\gamma$ exceeds $\dfrac{2\beta}{\sqrt{\alpha}}$, the equilibrium point becomes an unstable node losing its spiral behaviour. This is shown in figures 10, 11 and 12.

All the bifurcations discussed so far are local bifurcations which could be forecasted by linear stability analysis of the equilibrium points. Apart from these, the system also exhibits an important type of global bifurcation - the saddle homoclinic bifurcation.

 When $\gamma$ changes through $0.10$, the limit-cycle created in the Hopf bifurcation mentioned earlier eventually collides with the saddle creating a homoclinic orbit. If $\gamma$ increases further, the limit cycle disappear creating two unstable fixed points as shown in figures 13 and 14.

\begin{figure}[h]
\centering
  %\begin{center}
   %\includegraphics{Z_x_1.jpg}
 \includegraphics[trim=0cm 0cm 0cm 0cm, clip=true, height=7cm, width=0.5 \textwidth]{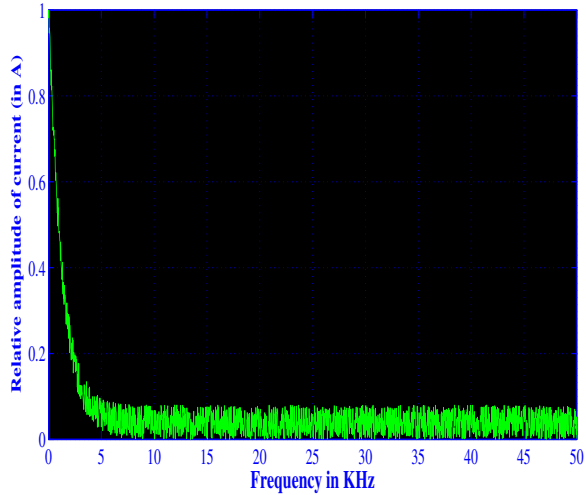}  
  \caption{Frequency response with a peak at zero frequency when $\gamma=0.18$}
 %\end{center}
\end{figure}

These homoclinic orbits have profound importance on the stability of a system and a lot of work had been devoted to study the behaviour near homoclinic orbit\cite{ref10}. The variables that exhibit sustained oscillations among the limit cycles vary exceeding slowly as the cycle approaches the homoclinic orbit. When the limit cycle disappear they increases in an unbounded manner. The homoclinic orbit represents an infinite-time oscillation, so its frequency response has a peak on zero frequency as shown in figure 15.

\section{Conclusion}

In this paper a simple circuit model is proposed which exhibits feature rich dynamics and a handful of bifurcations. The circuit model is easy to implement using passive elements and control sources and the bifurcations can be observed by varying linearly the control factor of a voltage-controlled current source within the realistic range of circuit parameters. Several local bifurcations and a global bifurcation are observed during circuit operation. The circuit can be useful for illustration purpose and experimental study of different bifurcations. The potential application of the circuit can be a controlled amplitude sustained oscillator where the limit-cyclic behaviour generated during the Hopf bifurcation can be exploited. It has the advantage that the oscillation will be stabled at a fixed amplitude that can be controlled by varying the parameters. This property is extremely useful in electronics industry.

All the circuit simulations accomplished using \emph{Linear Technology-Simulation Program with Integrated Circuit Emphasis(LTspice)}.

% references
%\section*{References}

\end{document}